\documentclass[aps,prd,twocolumn,superscriptaddress,nofootinbib,longbibliography,10pt]{revtex4-2}
\usepackage{etoolbox}
\usepackage{dcolumn}
\usepackage{amsmath,amssymb,amsfonts,mathtools,bm}
\usepackage{mathrsfs,bbold}
\usepackage{graphicx}
\usepackage[colorlinks=true,urlcolor=blue,citecolor=red,linkcolor=blue]{hyperref}
\usepackage{natbib}
\usepackage{wrapfig}
\usepackage{flushend}
\usepackage{BOONDOX-cal}

\begin{document}
\title{Quantum gravitational deflection of parallel matter wave beams}
\author{Soham Sen}
\email{sensohomhary@gmail.com}
\affiliation{Department of Astrophysics and High Energy Physics, S. N. Bose National Centre for Basic Sciences, JD Block, Sector-III, Salt Lake City, Kolkata-700 106, India}
\author{Vlatko Vedral}
\email{vlatko.vedral@physics.ox.ac.uk}
\affiliation{Clarendon Laboratory, University of Oxford, Park Road, Oxford OX1 3PU, United Kingdom}
\begin{abstract}
\noindent It is well known that two parallel photon beams do not deflect under the effect of their energy-momentum tensor. In this work, we propose a novel model where two spatially separated Bose-Einstein condensates are outcoupled to create two parallel atom laser beams. We find out that apart from the classical deflection, a purely quantum gravity induced tidal deflection is observed which results in an irreducible noise in the geodesic separation of the two beams. Based on this simple but novel theoretical outcome, we propose an experimental model for detecting this quantum gravity induced standard deviation in the geodesic separation of the two parallel matter-wave beams.
\end{abstract}
\maketitle
\section{Introduction}
\noindent The fundamental problem of a quantum field theory of gravity lies in the fact that it is non-renormalizable and the existing quantum gravity models run into serious intricacies. In recent times the main focus has therefore been shifted to detect low energy effects of quantum gravity using simple but intuitive experimental set-up. The main objective of low energy quantum gravity models is to look for signatures of quantum gravity at lower energy scales rather than directly probing the Planck length scale. It is already well known in the literature that the quanta of linearized gravity is a spin two particle known as a graviton \cite{SNGupta1,SNGupta2,Feynman,DeWitt,DeWitt2}.
Richard Feynman proposed a simple experiment to rectify whether quantum mechanics and gravity is compatible with each other or not using the concept of superposition of masses \cite{FeynmanQG,MarlettoVedralNature}. Based on this question raised an experiment was proposed which aims to investigate the gravity mediated generation of spin entanglement between two test masses which are in quantum superposition with each other \cite{Bose1,Bose2,Marletto}. These proposals have sparked the huge long debate of whether ``Is gravity really quantum?" and led to several literature in this direction \cite{Graviton_1,Graviton_2,Graviton_3,
Graviton_4,Graviton_5,Graviton_6,BEC_PRX}. These works are based purely on the proposal of quantum gravity induced entanglement of masses or QGEM protocol.  In a recent work \cite{SenVedral}, we have proposed a new proposal which is based on a completely new protocol known as the quantum gravity induced entanglement of phonons or QGEP protocol. We have, in this work, investigated the entanglement generation of phonons in two spatially separated static Bose-Einstein condensates. We have found out that the measure of entanglement strongly depends on the number of particles in the condensate is not only highly controllable in an experimental set-up but also results in significant entanglement generation. If the condensates are now not kept static rather the bosons are outcoupled from the magnetic traps they freely fall under the effect of Earth's gravitational field while the bosons remain coherent with each other behaving like a matter wave beam. This matter wave beam is known as an atom laser. Now, we discuss a very simple physical scenario.


\noindent If two parallel light beams travel while keeping parallel to each other in a flat background, they do not deflect as a result of the background perturbation generated by the energy momentum tensor. A very simple analysis using the geodesic equation depicting the trajectory of the photon beam is possible and we can write down the equation as
\begin{equation}\label{1.1}
\frac{d^2 x^\mu}{d\tau^2}+\Gamma^{\mu}_{~\alpha\beta}\frac{dx^\alpha}{d\tau}\frac{dx^\beta}{d\tau}=0~.
\end{equation}
From the linearized Einstein's equation, one can obtain the analytical expression for the metric fluctuation, in terms of the energy momentum tensor as
\begin{equation}\label{1.2}
h_{\mu\nu}=\frac{4G}{c^4}\int d^3\mathbf{r}'\frac{T_{\mu\nu}(t-|\mathbf{r}-\mathbf{r}'|,\mathbf{r}')}{|\mathbf{r}-\mathbf{r}'|}
\end{equation}
where the energy momentum tensor for photons $T_{\mu\nu}=\rho c^2 k_\mu k_\nu$ and the wave vector $k_\mu$ for a wave propagating in the $z$ direction reads $k_{\mu}=\{-1,0,0,1\}$. The wave perturbation $h_{\mu\nu}$ reads $h_{\mu\nu}=\Phi k_\mu k_\nu$ for $\rho(r)=\rho$ and $\Phi\equiv\frac{8\pi G\rho r^2 }{c^2}$.  One can then obtain the geodesic equation as $\ddot{x}=\frac{c^2}{2}\left(\partial_x\Phi+\partial_x\Phi-2\partial_x\Phi\right)=0$. This indicates a very simple physics that two parallel light beams never deflects under their own gravitational interaction. This is, however, not quite correct for phonons as they do not follow a null geodesic, that is $k_\mu k^\mu\neq 0$.  Based on this simple physical proposal, we propose a very simple experimental set-up for detecting graviton signatures in two parallel matter wave beams.
We at first revisit the analysis of the deflection of two atom laser beams under the effect generated due to the classical gravitational field as a result of the energy momentum tensor of the two macroscopic Bose-Einstein condensates \cite{SenVedral}. We have then treated the system quantum mechanically which results in an background perturbation which is operator-valued. We have then calculated the geodesic equation and eventually solved it to find whether really quantum gravitational irreducible noise is induced in the deflection of the trajectories of the atom-laser beams. Based on this simple but novel analytical model, we propose an experimental set-up which may be able to detect quantum gravitational deflection of the two parallel beams.
Our primary aim in this work is to discuss the classical model and then investigate the quantum gravitational analogue of the deflection of two parallel atom laser beams generated from two spatially separated Bose-Einstein condensates.
\section{The model}\label{S2}
\noindent We start by revisiting the deflection of the two parallel matter wave beams while they freely fall under the effect of the Earth's gravitational field from \cite{SenVedral} and in the next subsection, we move to its quantum analogue.
\subsection{The classical analogue}
\begin{figure}
\begin{center}
\includegraphics[scale=0.8]{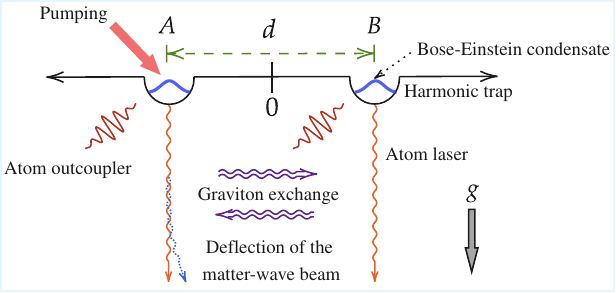}
\caption{Two Bose-Einstein condensates are generated inside of two harmonic traps such that they are separated by a finite spatial distance $d$. Using an atom outcoupler bosons are outcoupled from the magnetic trap and freely fall under the effect of Earth's gravitational field creating two freely falling atom laser beams.\label{Parallel_Atom_Laser_OTM}}
\end{center}
\end{figure}
\noindent We start by visualizing the model. A simple analogue of this model can be considered as given in Fig.(\ref{Parallel_Atom_Laser_OTM}), however, the figure really represents the quantum gravitational analogue. We find that two Bose-Einstein condensates are prepared in two magnetic traps at positions A and B while they are spatially separated by a distance $d$. The bosons are then outcoupled using atom outcoupler and radio frequencies while the quantum magnetic number in the state goes to zero resulting in the bosons to come out of the trap and fall freely under the effect of Earth's gravitational field. We calculate the geodesic equation for one of the atom laser beams in any spatial direction perpendicular to its direction of free-fall. As the atom-lasers are freely falling the contravariant and covariant wave vectors take the analytical expressions $k^{\mu}=\{1,0,0,\frac{v}{c}\}$ and $k_\mu=\{-1,0,0,\frac{v}{c}\}$. For a null trajectory $v=c$. The first step is to construct the energy momentum tensor for the system. For a Bose-Einstein condensate the ground state wave-function with non-interacting bosons can be obtained simply by solving the Gross-Pitaevski equation and the analytical expression for the normalized wave function is given by \cite{BEC_Pitaevski_Stingari}
\begin{equation}\label{1.3}
\psi(r)=\sqrt{N_0}\left(\frac{m\omega}{\pi\hbar}\right)^\frac{3}{4}e^{-\frac{m\omega r^2}{2\hbar}}
\end{equation}
with $\omega$ being the frequency, $m$ being the mass of each bosons and $N_0$ being the number of bosons in the ground state of the system. The energy momentum tensor can then takes the analytical form as \begin{equation}\label{1.4}
T_{\mu\nu}=m c^2 \psi^*(r)\psi(r) k_\mu k_\nu=N_0 mc^2 \left(\frac{m\omega}{\pi \hbar}\right)^\frac{3}{2}e^{-\frac{m\omega r^2}{\hbar}}k_\mu k_\nu
\end{equation}
where it is evident that the non-vanishing components of the energy momentum tensors are $T_{00}$, $T_{03}$, $T_{30}$, and $T_{33}$. We can now obtain the classical metric perturbation due to the energy momentum tensor of the model system simply by substituting the energy momentum tensor in eq.(\ref{1.2}) and then using it to construct the Christoffel connection, we can finally arrive at the geodesic equation in eq.(\ref{1.1}) as
\begin{equation}\label{1.5}
\frac{d^2 x^i}{dt^2}=8\pi G N_0 m x^i \left(\frac{m\omega}{\pi \hbar}\right)^\frac{3}{2}e^{-\frac{m\omega r^2}{\hbar}}\left(1-\frac{v^2}{c^2}\right)^2~.
\end{equation}
This is the same result also produced in our earlier work in \cite{SenVedral}. We now proceed towards constructing its quantum mechanical analogue. Instead of considering it as a macroscopic wave function, we consider small amplitude oscillations and treat the joint condensate system as a quantum matter source. As a result the energy momentum tensor becomes operator valued resulting in quantum mechanical perturbation of the spacetime as well as quantum gravitational fluctuations in the system and its trajectories.
\subsection{The quantum gravitational analogue of the deflection model}
\noindent We start by writing down the energy momentum tensor operator for the model system depicted in Fig.(\ref{Parallel_Atom_Laser_OTM}) which reads 
\begin{equation}\label{1.6}
\hat{T}_{\mu\nu}(\mathbf{r})=m c^2 \hat{\Psi}^\dagger(\mathbf{r})\hat{\Psi}(\mathbf{r})L^3(\delta(\mathbf{r}-\hat{\mathbf{r}}_A)+\delta(\mathbf{r}-\hat{\mathbf{r}}_B))k_\mu k_\nu
\end{equation}
with $L$ denoting the quantization length and the order parameter of the condensate with small amplitude oscillations reading 
\begin{equation}\label{1.7}
\hat{\Psi}(\mathbf{r})=\sqrt{N_0} \left(\frac{m\omega}{\pi\hbar}\right)^\frac{3}{4}e^{-\frac{m\omega r^2}{2\hbar}}(1+\varepsilon \hat{b}_{\mathbf{r}})
\end{equation}
where $\varepsilon\equiv \frac{1}{\sqrt{N_0}}\ll 1$ and $\hat{b}_r$ annihilates a quasi-particle (phonon) vacuum state at position $\mathbf{r}$. The operator-valued metric perturbation will then be given by the expression $\hat{h}_{\mu\nu}=\frac{4G}{c^4}\int d^3\mathbf{r}'\frac{\hat{T}_{\mu\nu}(t-|\mathbf{r}-\mathbf{r}'|,\mathbf{r}')}{|\mathbf{r}-\mathbf{r}'|}$ where for a consistent analytical expression the entire right hand side should be Weyl-ordered. For simplicity, we consider $\hat{\mathbf{r}}_A=\{\hat{x}_A,0,0\}$ and $\hat{\mathbf{r}}_B=\{\hat{x}_B,0,0\}$. The classical part of $\hat{x}_A$ and $\hat{x}_B$ will depend completely on where the origin is considered. As per Fig.(\ref{Parallel_Atom_Laser_OTM}), we can write down $\hat{x}_A$ and $\hat{x}_B$ as $\hat{x}_A=\frac{d}{2}+\sqrt{\frac{\hbar}{2m\omega}}(\hat{\alpha}+\hat{\alpha}^\dagger)$ and $\hat{x}_B=-\frac{d}{2}+\sqrt{\frac{\hbar}{2m\omega}}(\hat{\beta}+\hat{\beta}^\dagger)$\footnote{Here, $\hat{\alpha}\equiv \hat{b}_{\mathbf{r}_A}$ and $\hat{\beta}\equiv \hat{b}_{\mathbf{r}_B}$.}. The operator valued background fluctuation $\hat{h}_{\mu\nu}$
\begin{widetext}
\begin{equation}\label{1.8}
\begin{split}
\hat{h}_{\mu\nu}(t,x)=\frac{2N_0 G m}{c^2}\left(\frac{m\omega L^2}{\pi\hbar}\right)^\frac{3}{2}&\left[:|x-\hat{x}_A|^{-1}(1+\varepsilon \hat{\alpha}^\dagger)e^{-\frac{m\omega}{\hbar}(x-\hat{x}_A)^2}(1+\varepsilon \hat{\alpha}):_{W}\otimes\hat{\mathbb{1}}_B\right.\\&+\left.\hat{\mathbb{1}}_A\otimes|x-\hat{x}_B|^{-1}(1+\varepsilon \hat{\beta}^\dagger)e^{-\frac{m\omega}{\hbar}(x-\hat{x}_B)^2}(1+\varepsilon \hat{\beta}):_{W}\right]k_\mu k_\nu
\end{split}
\end{equation}
\end{widetext}
where $::_W$ denotes the symmetric or Weyl ordering of the operators. It is therefore now quite straightforward to write down the geodesic equation as
\begin{equation}\label{1.9}
\frac{d^2\hat{x}}{dt^2}=\frac{c^2}{2}\left[\partial_x \hat{h}_{00}+\frac{2v}{c}\partial_x \hat{h}_{03}+\frac{v^2}{c^2}\partial_x \hat{h}_{33}\right]
\end{equation}
and we can simply set $x$ to zero after taking the differential with respect to $x$ as it will not contribute towards any true quantum gravity effects while considering the equation up to the linear order in the quasi-particle raising and lowering operators. For a freely falling atom laser beams, it is evident that at each instant $v$ will change and it is the only component in the right hand side of eq.(\ref{1.9}) that is dependent on time. Our aim here is to calculate the deviation operator $\Delta\hat{x}=\hat{x}-\langle\hat{x}\rangle$ and then obtain the components for the square of the deviation operator. Here, the expectation is taken with respect to the ground state of the system $|\psi\rangle=|0_A,0_B\rangle$. After quite bit of analytical steps, we obtain the expression for the deviation operator up to first order in the quasi-particle ladder operators as\footnote{For a freely falling object the velocity at time $t$ is given by $v(t)=gt$ and, as a result, we obtain $\int_0^\tau dt\int_0^t dt' \left(1-\frac{v^2(t')}{c^2}\right)^2=\frac{\tau^2}{2}\left(1-\frac{g^2\tau^2}{3c^2}+\frac{g^4\tau^4}{15c^4}\right)\simeq \frac{\tau^2}{2}$ for significantly small free fall time.}
\begin{equation}\label{1.10}
\begin{split}
\Delta \hat{x}(\tau)\simeq&\frac{4N_0G m \tau^2}{d^2}\left[\frac{m\omega L^2}{\pi\hbar}\right]^\frac{3}{2}e^{-\frac{m\omega d^2}{4\hbar}}\\\times&\left[(\hat{\alpha}+\hat{\alpha}^\dagger)\mathcal{A}\otimes \hat{\mathbb{1}}_B+\hat{\mathbb{1}}_A\otimes \left(\hat{\beta}+\hat{\beta}^\dagger\right)\mathcal{B}\right]
\end{split}
\end{equation}
where the coefficients $\mathcal{A}$ and $\mathcal{B}$ are defined as
\begin{align}
\mathcal{A}&\equiv \left(1+\frac{m\omega d^2}{2\hbar}\right)\left(\varepsilon-\sqrt{\frac{m\omega d^2}{2\hbar}}\right)-2\sqrt{\frac{2\hbar}{m\omega d^2}}\label{1.11}\\
\mathcal{B}&\equiv \left(1+\frac{m\omega d^2}{2\hbar}\right)\left(\varepsilon+\sqrt{\frac{m\omega d^2}{2\hbar}}\right)+2\sqrt{\frac{2\hbar}{m\omega d^2}}~.
\end{align}
For $d\gtrsim \sqrt{\frac{2\hbar}{m\omega}}$, both $\mathcal{A}$ and $\mathcal{B}$ takes a very simple structure given as $\mathcal{B}=-\mathcal{A}\simeq \left(\frac{m\omega d^2}{2\hbar}\right)^\frac{3}{2}$. One can then obtain the standard deviation in the $x$ direction by calculating $\Delta x=\sqrt{\langle\psi|(\Delta \hat{x}(\tau))^2|\psi\rangle}$ where $|\psi\rangle$ denotes the ground state of the system. The analytical expression for $\Delta x$ then reads
\begin{equation}\label{1.13}
\Delta x(\tau)\sim \frac{4 \sqrt{2}N_0 G m \tau^2}{d^2}\left(\frac{m\omega d^2}{2 \hbar}\right)^\frac{3}{2}e^{-\frac{m\omega d^2}{4\hbar}}
\end{equation}
where we have considered the quantization length to be of the order of $L\sim \sqrt{\frac{\pi\hbar}{m\omega}}$ and $d\gtrsim \sqrt{\frac{2\hbar}{m\omega}}$. If one now set the parameter values to $N_0=10^6$, $\omega=10^3$ Hz, $m=10^{-25}$ kg and $d\simeq 2\sqrt{\frac{2\hbar}{m\omega}}$, we can obtain the standard deviation in $x$ to be $\Delta x\sim 10^{-18}\tau^2$ m. For atom lasers the laboratory scale free fall time is of the order of $\tau\sim 10^{-1}$ sec and can have a value as high as 1 sec, and thus $\Delta x\sim 10^{-18}-10^{-21}$ m. This can be controlled by changing the number of the particles in the BEC. The important thing to remember is that this deflection is purely irreducible quantum noise and as a result of which we are observing this non-vanishing standard deviation. Now, for a more physical scenario, it is more prudent to understand this effect as a scenario where the vacuum states of the two system undergoes a graviton exchange scenario, as a result of which the atom laser beams deflects while getting induced by an irreducible quantum noise. For $d\lesssim \sqrt{\frac{2\hbar}{m\omega}}$, the standard deviation takes the simple form
\begin{equation}\label{1.14}
\Delta x(\tau)\sim \frac{16 N_0 G\tau^2}{d^3}\sqrt{\frac{m\hbar}{\omega}}e^{-\frac{m\omega d^2}{4\hbar}}\sim\frac{16 N_0 G\tau^2}{d^3}\sqrt{\frac{m\hbar}{\omega}}
\end{equation}
which shows a $\frac{1}{d^3}$ like behaviour for very close spatial separation between the condensates.
Based on the above analytical outcomes, we shall now propose a very simple but intuitive experiment to detect any quantum gravitational signatures in the deflection of two parallel atom-laser beams.

 \subsection{Experimental proposal}
\noindent We start by considering a cavity-QED set up where two Bose-Einstein condensates are prepared in spatially separated magnetic harmonic traps. The bosons are then outcoupled using atom outcouplers from the magnetic traps and then they freely fall under the effect of the Earth's gravitational field while the bosons remain coherent to each other. We now consider two identical set ups of spatially separated condensates with pumping and atomic outcouplers. We consider that the number of bosons in the condensates of ``Set 1" $N_0^{(1)}$ is several orders of magnitude larger than the number of boson $N_0^{(2)}$ in ``Set 2".
\begin{widetext}
\begin{center}
\begin{figure}[ht!]
\includegraphics[scale=0.825]{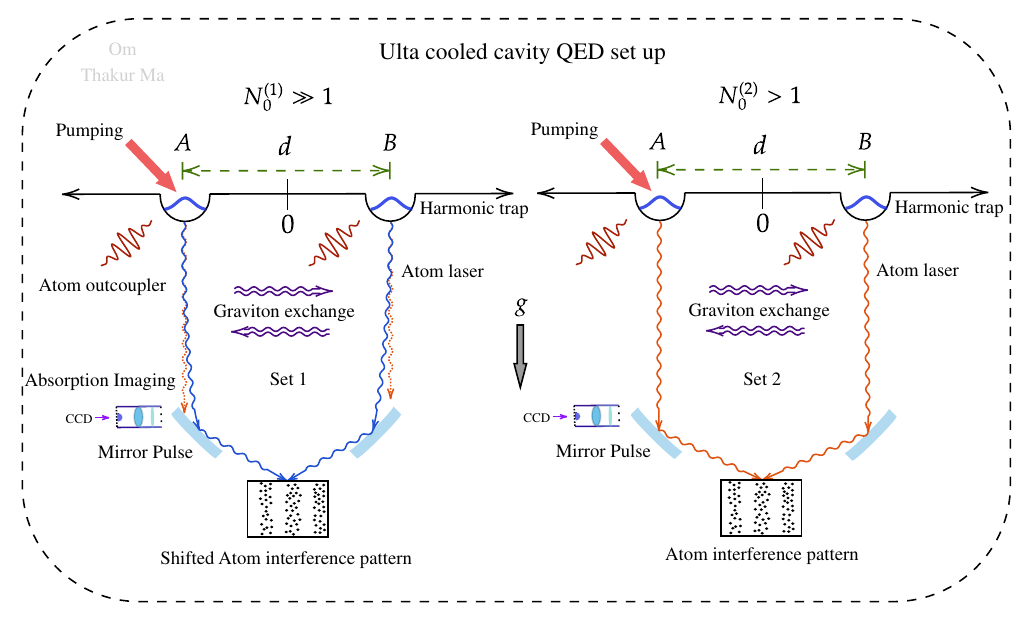}
\caption{Two identical matter wave interferometers are placed inside a cavity QED set-up where ``Set 1" uses condensates with each having $N_0^{(1)}$ number of bosons such that $N_0^{(1)}\gg N_0^{(2)}$ with $N_0^{(2)}$ denoting the number of bosons in ``Set 2".\label{Detector_Om_Thakur_Ma}}
\end{figure}
\end{center}
\end{widetext}
In  ``Set 1", the orange dotted lines denote the undeflected paths of the atom laser beams whereas the blue curvy lines denote the deflected paths. After a travel time of $\tau\sim 10^{-1}$ sec, the beams are then interfered using two mirror pulses to create a matter wave interference pattern. As the standard deviation $\Delta x$ is proportional to the number of particles in the condensate systems  (as can be seen from eq.(s)(\ref{1.13},\ref{1.14})), the deflection becomes negligible for small value of $N_0$. In ``Set 2", we have used the similar set up but with condensates with a significantly lower number of particles such that $N_0^{(1)}\gg N_0^{(2)}$. It is therefore evident that $\Delta x^{(2)}\ll \Delta x^{(1)}$. The standard deviation will result in a phase difference and an eventual shift in the matter wave interference pattern and a side by side comparison will reveal a significant patter shift in ``Set 1" compared to ``Set 2". A repeated analysis then shall help one to determine the standard deviation due to the graviton induced irreducible noise in the geodesic direction $\Delta x$. This indeed is a very interesting outcome and if detected will result in detection of quantum gravity induced deflection of atom laser beams.

\noindent \textbf{\textit{An estimate of the  fringe shift and the possibility of detection}:} We shall now provide an estimate for fringe shift in the atom interference pattern due to the standard deviation $\Delta x$. The local standard deviation in the phase shift in the atom laser beams can be calculate simply by $\Delta\theta=k\Delta x$. The first step is to calculate $k$ which is related to the velocity and frequency of the phonons in the condensate as $k=\frac{2\pi}{\lambda}=\frac{\omega}{v}$. Although we are considering non-interacting Bose gases in a realistic scenario $g$ will be non-vanishing $g=\frac{4\pi\hbar^2a}{m}$. Consider for $\mathrm{Rb}_{87}$ the scattering length is $a=100-105 ~a_0\simeq 5.29\times 10^{-9}$ m with $a_0$ denoting the Bohr radius. This results in the interaction parameter $g$ having a value of $g\sim 10^{-50}$ $\mathrm{J}\cdot\mathrm{m}^{3}$. The velocity of the phonons is given by $v=\sqrt{\frac{gn}{m}}\sim 5.5\times 10^{-2}$ $\mathrm{m}\cdot\mathrm{sec}^{-1}$ with $n\sim \frac{N_0}{V}\sim 4. \times 10^{22}$ $\mathrm{m}^{-3}$. For $\omega=10^3$ Hz then one can obtain $\Delta\theta$ to be $\Delta \theta= k\Delta x\sim 10^{-13}$ $\mathrm{rad}$. This results in an overall fringe shift $\Delta f=\frac{\Delta\theta}{2\pi}\times \ss$ with $\ss$ denoting the fringe width in the atom interference pattern. In standard atom interferometers the fringe width is practically of the order of few hundred nanometers, however, in high precision experimental set ups, it is possible to obtain a read out $\ss$ of the order of $\ss\sim 10^{-4}$ m. This results in a fringe shift of the order of $\Delta f\sim 10^{-18}$ m. Current atom interferometers reach a maximum precision of $10^{-6}$ rad which is equivalent to $10^{-10}$ m of fringe shift. The first step to increase the phase shift is by increasing $N_0\sim 10^{7}-10^{9}$ (magnon BEC) and increase the free fall time up to $\tau\sim 10$ sec. This results in an overall increase of $10^3-10^5$ in the fringe shift resulting in $\Delta f \sim 10^{-13}-10^{-15}$ m. Using LMTs or Large momentum transfer one can also amplify the phase gain by an order of $10^3$ resulting in an final effective phase shift of $\Delta \theta\sim 10^{-5}$ rad which is in line with the standard quantum limit and $\Delta f\sim10^{-10}$ m.

\section{Conclusion}\label{S3}
\noindent In this work, we investigate a novel set up where two freely falling atom laser beams from two spatially separated Bose-Einstein condensates get deflected under the perturbation created in the background by the interaction of quantum matter. We revisit the basic case of parallel photon beam deflection and move towards the phonon-phonon deflection case. We consider the simple geodesic equation where we calculate the deviation in the direction perpendicular to the propagation direction of the matter wave beams. We find out that considering the background fluctuation to be operator valued as a result of the back reaction of quantum matter on the spacetime, the standard deviation in the direction perpendicular to the propagation vector becomes non-vanishing. This phenomena indicates that the two parallel matter wave beams get deflected while the geodesic get affected by irreducible quantum gravitational noise as a result of quantum gravity induce vacuum fluctuation and we find out a maximum value for this deflection to be of the order of $10^{-13}$ m for a BEC system with $N_0\sim 10^6$ number of atoms. We have then also investigated the result in the delocalized wave packet or low $d$ regime where we find out that the standard deviation is proportional to $d^{-3}$ leading to a tidal Newton-like deflection effect. Finally, we have proposed an experimental model based on our analytical results and have provided an estimate for the effective fringe shift in our proposed model and the effective way to enhance the phase accumulation and eventually the phase shift. We hope that if implemented the experimental set-up will actually help in detecting quantum gravitational deflection of phonons.


\begin{thebibliography}{20}
\bibitem{SNGupta1}
S. N. Gupta, ``\textit{Quantization of Einstein's Gravitational Field: Linear Approximation}", \href{https://iopscience.iop.org/article/10.1088/0370-1298/65/3/301}{Proc. Phys. Soc. A 65 (1952) 161}.
\bibitem{SNGupta2}
S. N. Gupta, ``\textit{Gravitation and Electromagnetism}", \href{https://doi.org/10.1103/PhysRev.96.1683}{Phys. Rev. 96 (1954) 1683}.
\bibitem{Feynman}
R. P. Feynman, ``\textit{Quantum theory of gravitation}", \href{https://www.researchgate.net/profile/Valery-Morozov/post/Does_spacetime_possess_the_properties_of_a_relativistic_aether/attachment/60db66806160740001e6a69d/AS%3A1040103304601607%401624991360735/download/Feynman_1963.pdf}{Acta. Phys. Pol. 24 (1963) 697}.
\bibitem{DeWitt}
B. S. DeWitt, ``\textit{Quantum Theory of Gravity. I. The Canonical Theory}", \href{https://doi.org/10.1103/PhysRev.160.1113}{Phys. Rev. 160 (1967) 1113}. 
\bibitem{DeWitt2}
B. S. DeWitt, ``\textit{Quantum Theory if Gravity. II. The Manifestly Covariant Theory}", \href{https://doi.org/10.1103/PhysRev.162.1195}{Phys. Rev. 162 (1967) 1195}.
\bibitem{FeynmanQG}
R. P. Feynman  in The Role of Gravitation in Physics: \href{https://edition-open-sources.org/media/sources/5/Sources5.pdf}{Report from the 1957 Chapel Hill Conference} (eds D. Rickles, C. M. DeWitt) (Edition Open Sources, 2011).
\bibitem{MarlettoVedralNature}
C. Marletto and V. Vedral, ``\textit{Witness gravity's quantum side in the lab}", \href{https://www.nature.com/articles/547156a}{Nature 547 (2017) 156}.
\bibitem{Donoghue}
J. F. Donoghue, ``\textit{General relativity as an effective field theory: The leading quantum corrections}", \href{https://doi.org/10.1103/PhysRevD.50.3874}{Phys. Rev. D 50 (1994) 3874}.
\bibitem{Bose1}
S. Bose, A. Mazumdar, G. W. Morley, H. Ulbricht, M. Toro\v{s}, M. Paternostro, A. Geraci, P. Barker, M. S. Kim, and G. Milburn, ``\textit{Spin Entanglement Witness for Quantum Gravity}", \href{https://journals.aps.org/prl/abstract/10.1103/PhysRevLett.119.240401}{Phys. Rev. Lett. 119 (2017) 240401}.
\bibitem{Bose2}
R. J. Marshman, A. Mazumdar, and S. Bose, ``\textit{Locality and entanglement in table-top testing of the quantum nature of linearized gravity}", \href{https://journals.aps.org/pra/abstract/10.1103/PhysRevA.101.052110}{Phys. Rev. A 101 (2020) 052110}.
\bibitem{Marletto}
C. Marletto and V. Vedral, ``\textit{Gravitationally Induced Entanglement between Two Massive Particles is Sufficient Evidence of Quantum Effects in Gravity}", \href{https://journals.aps.org/prl/abstract/10.1103/PhysRevLett.119.240402}{Phys. Rev. Lett. 119 (2017) 240402}.
\bibitem{Graviton_1}
D. L. Danielson, G. Satischandran, and R. M. Wald, ``\textit{Gravitationally mediated entanglement: Newtonian field vs gravitons}", \href{}{Phys. Rev. D 105 (2022) 086001}.
\bibitem{Graviton_2}
J. S. Pedernales, G. W. Morley, and M. B. Plenio, ``\textit{Motional Dynamical Decoupling for Matter-Wave Interferometry}", \href{https://doi.org/10.1103/PhysRevLett.125.023602}{Phys. Rev. Lett. 125 (2020) 023602}. 
\bibitem{Graviton_3}
R. J. Marshman, A. Mazumdar, R. Folman, and S. Bose, ``\textit{Large splitting massive Schr\"{o}dinger kittens}", \href{https://doi.org/10.1103/PhysRevResearch.4.023087}{Phys. Rev. Research 4 (2022) 023087}.
\bibitem{Graviton_4}
M. Christodoulou and C. Rovelli, ``\textit{On the possibility of laboratory evidence for quantum superposition of geometries}", \href{https://doi.org/10.1016/j.physletb.2019.03.015}{Phys. Lett. B 792 (2019) 64}. 

\bibitem{Graviton_5}
D. Mikki, A. Matsumura, and K. Yamamoto, ``\textit{Entanglement and decoherence of massive particles due to gravity}", \href{https://doi.org/10.1103/PhysRevD.103.026017}{Phys. Rev. D 103 (2021) 026017}.
\bibitem{Graviton_6}
A. Matsumara and K. Yamamoto, ``\textit{Gravity-induced entanglement in optomechanical systems}", \href{https://doi.org/10.1103/PhysRevD.102.106021}{Phys. Rev. D 102 (2020) 106021}.
\bibitem{BEC_PRX}
R. Howl, V. Vedral, D. Naik, M. Christodoulou, C. Rovelli, and A. Iyer, ``\textit{Non-Gaussianity as a Signature of a Quantum Theory of Gravity}", \href{https://doi.org/10.1103/PRXQuantum.2.010325}{PRX Quantum 2 (2021) 010325}.
\bibitem{SenVedral}
S. Sen, S. Gangopadhyay, and V. Vedral, ``\textit{Gravity mediated entanglement of phonons in Bose-Einstein condensates}", \href{https://doi.org/10.48550/arXiv.2604.20767}{arXiv:2604.20767 [hep-th]}.
\bibitem{BEC_Pitaevski_Stingari}
L. Pitaevski and S. Stingari, ``\textit{Bose-Einstein Consensation and Superfluidity}" (Oxford University Press, Oxford, 2016).



\end{thebibliography}
\end{document}